\documentclass[aps,prb,twocolumn,superscriptaddress]{revtex4}

\input epsf.sty

\begin{document}

\preprint{8/4/02 ver16}

%
%

\title{Mode-coupling and polar nanoregions in the
relaxor ferroelectric Pb(Mg$_{1/3}$Nb$_{2/3}$)O$_{3}$}

\author{S. Wakimoto}
\email[Corresponding author: ]{waki@physics.utoronto.ca}
\affiliation{Department of Physics, Brookhaven National Laboratory,
  Upton, NY 11973}
\affiliation{Department of Physics, University of Toronto, Toronto,
  Ontario, Canada M5S~1A7}

\author{C. Stock}
\affiliation{Department of Physics, University of Toronto, Toronto,
  Ontario, Canada M5S~1A7}

\author{Z.-G. Ye}
\affiliation{Department of Chemistry, Simon Fraser University,
  Burnaby, British Columbia, Canada V5A~1S6}

\author{W. Chen}
\affiliation{Department of Chemistry, Simon Fraser University,
  Burnaby, British Columbia, Canada V5A~1S6}

\author{P. M. Gehring}
\affiliation{NIST Center for Neutron Research, National Institute of
  Standards and Technology, Gaithersburg, MD 20899}

\author{G. Shirane}
\affiliation{Department of Physics, Brookhaven National Laboratory,
  Upton, NY 11973}

\date{\today}

\begin{abstract}

We present a quantitative analysis of the phonon lineshapes
obtained by neutron inelastic scattering methods in the relaxor
ferroelectric Pb(Mg$_{1/3}$Nb$_{2/3}$)O$_3$ (PMN).  Differences in
the shapes and apparent positions of the transverse acoustic (TA)
and transverse optic (TO) phonon peaks measured in the (300) and
(200) Brillouin zones at 690\,K are well described by a simple
model that couples the TA and soft TO modes in which the primary
parameter is the wave vector and temperature-dependent TO
linewidth $\Gamma(q,T)$.  This mode-coupling picture provides a
natural explanation for the uniform displacements of the polar
nanoregions, discovered by Hirota {\it et al.}, as the PNR result
from the condensation of a soft TO mode that also contains a large
acoustic component.

\end{abstract}


\maketitle

\section{Introduction}

Considerable advances have been made over the past several years
in our understanding of the lattice dynamics of the relaxor
ferroelectric Pb(Mg$_{1/3}$Nb$_{2/3}$)O$_3$ (PMN), which is a
prototypical member of a class of compounds that possesses
exceptional piezoelectric properties, and an unusually broad and
frequency-dependent dielectric susceptibility that peaks at a
temperature $T_{\rm max}$ ($\approx 265$\,K at 1\,kHz for
PMN).~\cite{Ye_review}  Pioneering neutron scattering work on PMN
in 1999 by Naberezhnov {\it et al}.~\cite{Naberezhnov_99} has been
followed by a systematic series of neutron scattering experiments
by Gehring {\it et
al.}~\cite{Gehring_00_1,Gehring_proc_00,Gehring_00_2,Gehring_01}
and Wakimoto {\it et al.}~\cite{waki_02}  The results of these
latter measurements have firmly established the existence of
significant soft mode dynamics in PMN.  In particular, they have
demonstrated a linear temperature dependence of the zone-center
soft phonon energy squared $(\hbar\omega_{0})^2$ over a wide
temperature range, as shown in Fig.~1.

A fundamental feature that appears common to all relaxors is that
of the polar nanoregions, or PNR.  The presence of these regions
was inferred in 1983 by Burns and Dacol, who found that the optic
index of refraction of PMN deviates from a linear temperature
dependence at a temperature $T_d \approx 620$\,K that is far above
$T_{\rm max}$.~\cite{Burns_83}  In conventional ferroelectrics
this deviation corresponds to a uniform polarization that only
develops below $T_c$.  Given the absence of any net polarization
in PMN at $T_d$, the optical data indicate the formation of tiny
regions of local and randomly-oriented polarization that were
speculated to originate in Nb-rich parts of the crystal several
unit cells in size.  The dynamical effects of these PNR are
clearly manifest in neutron scattering measurements.  Gehring {\it
et al}.\ have found that the long-wavelength transverse optic (TO)
phonons are heavily damped below $T_d$, which gives rise to the
anomalous ``waterfall'' feature.~\cite{Gehring_01} Neutron
scattering measurements further suggest that the appearance of the
PNR at $T_d$ is accompanied by a remarkable overdamping of the
zone center ($q = 0$) TO phonons. As shown in Fig.~1(a), the zone
center TO phonon linewidth broadens with decreasing temperature
until finally at $T_{d}$ no discernible phonon peak remains. It
has subsequently been reported that the overdamped zone-center TO
mode reappears below $\sim 220$\,K.~\cite{waki_02}

The formation of the PNR at $T_d$ is the most important aspect of
the relaxor problem, and it is intriguing that it is accompanied
not only by the overdamping of the zone-center TO soft mode, but
also by the appearance of diffuse scattering.  The Brillouin zone
dependence of the neutron diffuse scattering intensities has been
extensively measured by Vakhrushev {\it et
al.},~\cite{Vakhrushev_89,Vakhrushev_93} and later it was shown by
Naberezhnov {\it et al}.\ that the onset of the diffuse scattering
occurs at $T_d$.~\cite{Naberezhnov_99}  The coincidence of these
phenomena at $T_d$ point to a picture in which the PNR result from
the condensation of the soft TO mode.  If this picture is correct,
then the atomic displacements associated the diffuse scattering
must match those associated with the soft TO phonon, namely they
must satisfy the center-of-mass condition.  However, the diffuse
scattering intensities in different Brillouin zones are entirely
inconsistent with the TO phonon intensities.  This long-standing
puzzle was resolved recently by Hirota {\it et
al}.~\cite{Hirota_01}  By calculating the dynamic structure factor
with the assumption that the soft TO mode condenses with an
additional uniform shift of all atoms, they demonstrated that the
intensities of TO phonon and diffuse scattering can be completely
reconciled.  Their calculation shows that this uniform shift
$\delta_{shift}$ is comparable to the displacement due to the TO
mode condensation $\delta_{c.m.}$, and that $\delta_{shift}$ is
60~\% of the displacement of Pb atom.  While the origin of this
uniform phase shift is unknown, the success of the model provides
further support to the idea connecting the PNR and the soft mode.

In this paper we demonstrate that a simple model that couples the
transverse acoustic and soft optic modes, similar to that used by
Harada {\it et al}.\ to describe the asymmetric phonon lineshapes
found in BaTiO$_3$,~\cite{Harada_71} is able to reproduce the
phonon lineshapes observed in PMN in both the (200) and (300)
Brillouin zones, which exhibit very different scattering profiles.
This naturally implies that a condensation at $T_d$ of a {\it
coupled} TO mode would then contain a TA component, and this might
closely relate to the uniform phase shift reported by Hirota {\it
et al.}~\cite{Hirota_01}

\begin{figure}
\centerline{\epsfxsize=3in\epsfbox{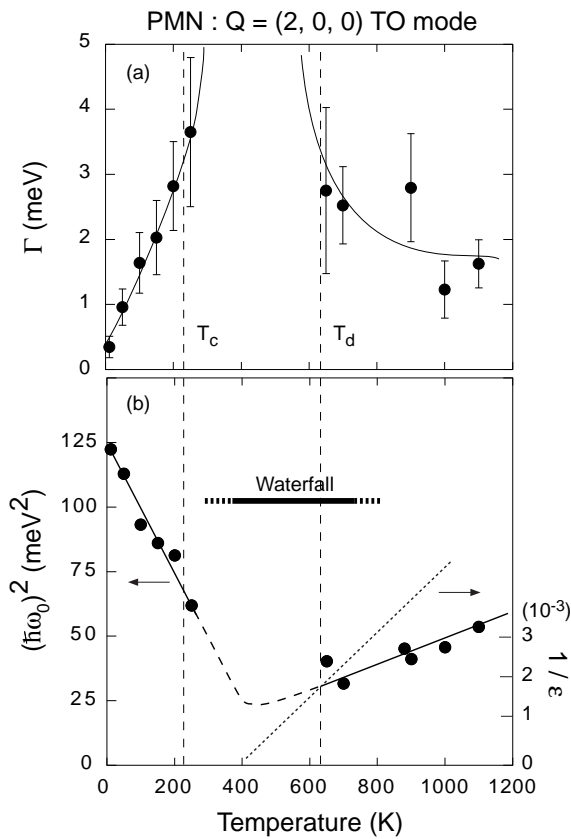}}
\caption{(a)Temperature dependence of the $(2,0,0)$ zone center TO
phonon linewidth $\Gamma_1$.  Above $T_{d}$ $\Gamma_1$ increases
gradually with decreasing temperature, whereas below $T_{c}$ it
decreases rapidly.  For temperatures between $T_{c}$ and $T_{d}$
it is impossible to determine the linewidth due to the overdamped
nature of the TO mode.  Solid lines are guides to the eye. (b)
Temperature dependence of the $(2,0,0)$ zone center TO phonon
energy squared $(\hbar\omega_{0})^{2}$ from Ref.~7. The dotted
line shows the Curie-Weiss behavior of the dielectric constant
$1/\epsilon \propto T-T_{0}$ with $T_{0} = 400$\,K reported by
Viehland {\it et al.}}
\end{figure}

\section{Experimental details}

Single crystals of PMN were grown using a top-seeded solution
method with PbO as flux.  Two crystals were chosen for the neutron
scattering experiments.  One has a volume of 0.09\,cm$^{3}$ and a
mass of 0.74\,g, and the other has a volume of 0.40\,cm$^{3}$ and
a mass of 3.25\,g (the density of PMN is 8.13\,gm/cm$^3$).  Since
both samples show quantitatively similar results, we will not
distinguish between the two samples in this paper.

Our neutron scattering experiments were performed at the BT9
triple-axis spectrometer located at the NIST Center for Neutron
Research.  The (002) reflection of highly-oriented pyrolytic
graphite (HOPG) crystals was used to monochromate the incident neutron
energy $E_{i}$ and to analyze final neutron energy $E_{f}$.  The
data were taken with a fixed final energy $E_{f} = 14.7$\,meV
($\lambda = 2.36$\,\AA$^{1}$), and horizontal collimations of
40$'$-40$'$-S-40$'$-80$'$ (``S"=sample) between source and
detector. An HOPG filter was placed after the analyzer to
eliminate higher order wavelength contamination in the scattered
beam.  The crystal was placed with a natural \{100\}$_{cubic}$
facet facing down on to a boron nitride post, and held in place
with tantalum wire. This orientation allows access to $(h0l)$-type
reflections in the horizontal scattering plane.  The mosaic
spreads of the crystals were less than 24$'$ at the (200)
reflection which is resolution-limited with the above configuration,
and indicate a high crystal quality.  The room temperature lattice 
constant is 4.04\,\AA, therefore one reciprocal lattice unit (rlu) 
equals 1.553\,\AA$^{-1}$.

\begin{figure}
\centerline{\epsfxsize=3.4in\epsfbox{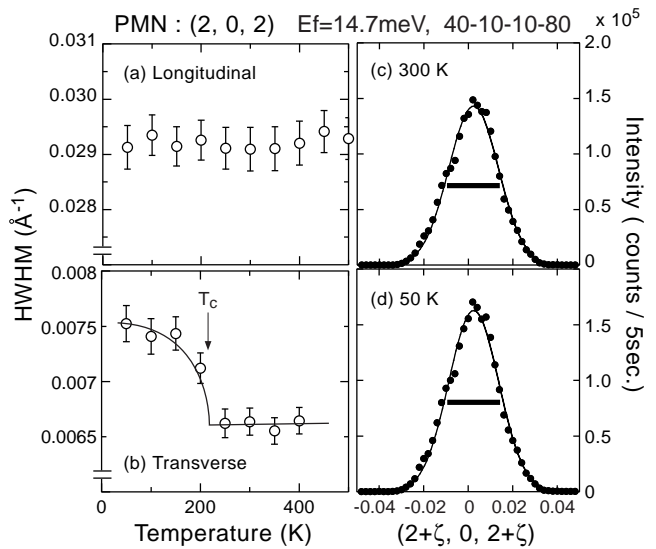}}
\caption{Temperature dependence of the $(2, 0, 2)$ Bragg peak
linewidth measured along the (a) $[101]$ (longitudinal), and (b)
$[10\bar{1}]$ (transverse) directions.  If the system transforms
from a cubic to a rhombohedral phase, then the $(2, 0, 2)$ peak
should split along the longitudinal direction.  Instead, only the
transverse scan shows a clear change near $T_{c}$.  For ease of
comparison, panels (c) and (d) show the $(2, 0, 2)$ lineshape
measured along the longitudinal direction at 300\,K and 50\,K. The
solid bars represent the instrumental longitudinal
$q$-resolution.}
\end{figure}

PMN reportedly remains cubic down to 5\,K.~\cite{deMathan}
Consistent with this observation, we find no definitive evidence
of a structural transition in the temperature range $50 \leq T
\leq 400$\,K.  On the other hand, a macroscopic ferroelectric
state can be established in PMN by cooling it in a moderate ($E =
1.7$\,kV/cm) applied electric field.  The induced polarization
then vanishes upon warming above $T_c = 213$\,K as a first-order
phase transition.~\cite{Ye_review} To date, the aforementioned
reappearance of the zone-center TO mode, which also coincides with
the disappearance of the TA phonon broadening,~\cite{waki_02} is
the most direct evidence of ferroelectric ordering in zero field
associated with $T_c$.  Other anomalies have also been reported
near $T_c$, such as the sharp peak in the temperature dependence
of the hypersonic damping reported by Tu {\it et
al}.,~\cite{Tu_95} and the abrupt change in the thermal expansion
of the lattice parameter observed by Dkhil {\it et
al.}~\cite{Dkhil_02}  To pursue this issue more carefully, we
looked for subtle structural changes near $T_c$ by employing a
tighter sequence of horizontal collimations
(40$'$-10$'$-S-10$'$-80$'$) to improve the instrumental $q$
resolution.  Figure 2 shows the temperature dependence of the
$(2,0,2)$ Bragg peak width along the (a) $[101]$ (longitudinal)
and (b) $[10\bar{1}]$ (transverse) directions.  If PMN were to
transform from a cubic to a rhombohedral structure, then the
$(2,0,2)$ Bragg peak would split along the longitudinal direction.
However, we observed neither a splitting nor a broadening of the
$(2,0,2)$ Bragg peak along $[101]$.  Figures 2(c) and (d) show
longitudinal scans of the $(2,0,2)$ Bragg peak measured at 300\,K
and 50\,K.  The peak width is almost resolution limited for $50
\leq T \leq 300$\,K. Instead, we observed a small, but clear, jump
in the transverse width at $T_{c}$ which is shown in Fig.~2(b).
However, we did not observe a corresponding enhancement of the
peak intensity below $T_c$ that is normally expected at a
structural phase transition due to the release of primary
extinction.

\section{Mode coupling model cross section}

The neutron inelastic scattering technique provides a direct
measure of the scattering function $S({\rm\bf q},\omega)$, as this
is simply related to the experimentally measured scattering cross
section.  More importantly, $S({\rm\bf q},\omega)$ is related to
the imaginary part of the dynamic susceptibility $\chi''({\rm\bf
q},\omega)$ via the fluctuation-dissipation theorem
\begin{equation}
S({\rm\bf q},\omega) = \{1+n(\omega)\}\chi''({\rm\bf q},\omega)\,
,
\end{equation}
where $n(\omega)$ is the Bose factor $(e^{\omega/k_{B}T}-1)^{-1}$.
By choosing the scattering vector ${\rm\bf Q} = {\rm\bf G} +
{\rm\bf q}$, where ${\rm\bf G}$ is a reciprocal lattice vector and
${\rm\bf q}$ is a phonon wave vector, one can measure the phonon
dispersions in different Brillouin zones.  The measurements
reported here were made at ${\rm\bf G}=(2,0,0)$ and $(3,0,0)$.
However a model cross section is required in order to extract
meaningful parameters from the measurements, and to correct for
the effects of the instrumental resolution. The form we chose for
$S({\rm\bf q},\omega)$ is discussed by Harada {\it et
al}.,~\cite{Harada_71} and pertains to a system in which a
coupling exists between two vibrational modes. In this case we
have
\begin{equation}
\begin{array}{cc}
S({\rm\bf q},\omega) = \{1+n(\omega)\}
         \frac{\omega}{A^2 + \omega^2 B^2}
         \{[(\Omega_{2}^{2} - \omega^{2})B - \Gamma_{2} A]F_{1}^{2}\\ \\
                  + 2 \lambda B F_{1}F_{2}
                  + [(\Omega_{1}^{2} - \omega^{2})B - \Gamma_{1}A]F_{2}^{2}\},
\end{array}
\end{equation}
where
\begin{equation}
A = (\Omega_{1}^{2} - \omega^{2})(\Omega_{2}^{2} - \omega^{2})
    - \omega^2 \Gamma_{1} \Gamma_{2} - \lambda^{2},
\end{equation}
\begin{equation}
B = \Gamma_{1}(\Omega_{2}^{2} - \omega^{2})
    + \Gamma_{2}(\Omega_{1}^{2} - \omega^{2}).
\end{equation}
The model parameters $\Omega$, $\Gamma$, $F$, and $\lambda$
represent the phonon frequency, damping, dynamic structure factor,
and coupling constant, while the indices 1 and 2 denote the TO and
TA modes, respectively.  Hereafter, we shall refer to this as the
mode-coupling (MC) function. This function shows larger coupling
effects the closer the TO and TA frequencies become, or the more
heavily damped either mode becomes. In the appendix we identify
several important features of the MC function that can produce
large differences in the phonon profiles.

\section{Neutron scattering experimental results}

\subsection{Phonon profiles at 690 K}

Substantial differences between the phonon profiles obtained in
the $(2,0,q)$ and $(3,0,q)$ Brillouin zones are apparent in
Fig.~3, which displays typical data sets for $q = -0.2$ measured
at 690\,K.  A major difference is that the scattering associated
with the TA phonon appears to peak at different energies, i.\ e.\
the TA energy seems lower at $(3,0,-0.2)$ than it is at
$(2,0,-0.2)$.  This difference is systematically observed at
different $q$ positions as shown in the inset of Fig. 3, where the
apparent TA phonon energies obtained from the peak position are
plotted as a function of $q$ in the (200) and (300) zones at
690\,K.  The solid lines are merely guides to the eye.  As is
clearly seen, the TA mode energies in the (300) zone are
consistently lower than those in (200) at all values of $q$
studied.  Of course, for a crystal that possesses true long-range
translational order, all phonon energies in different Brillouin
zones must be identical.

When fitting these phonon profiles with the MC function, we found
that a change in the sign of the product of the dynamic structure
factors $F_{1}F_{2}$ can reconcile the discrepancies in the TA
phonon peak position.  Specifically, if we set $F_1F_2 < 0$ for
$(2,0,-0.2)$ and $F_1F_2 > 0$ for $(3,0,-0.2)$, we obtain the same
TA phonon energy $\Omega_2$ for each zone.  This occurs because
for positive $F_{1}F_{2}$ the spectral weight of the TO mode
increasingly shifts to the region below the TA mode energy
$\Omega_2$ with increasing TO mode damping $\Gamma_1$, whereas for
negative $F_1F_2$ this spectral weight shifts to the region
between the TO and TA mode energies.  Thus the MC function
produces a higher (lower) apparent TA phonon energy for negative
(positive) $F_{1}F_{2}$, even though $\Omega_{2}$ is always the
same. Further details are given in the Appendix.
The fits to the MC function include an additional Gaussian
component which is used to describe the elastic incoherent peak at
$\hbar\omega = 0$. The fitted curves are shown in Fig.~3, where
the solid and dashed lines correspond to the MC and Gaussian
functions, respectively.  The fitting parameters are listed in
Table~1. It is satisfying to note that the values for $\Omega$ and
$\Gamma$ derived from the fits of both the $(2,0,-0.2)$ and $(3,
0, -0.2)$ phonon profiles agree quite well, even though the
apparent TA peak positions are different.

\begin{figure}
\centerline{\epsfxsize=3in\epsfbox{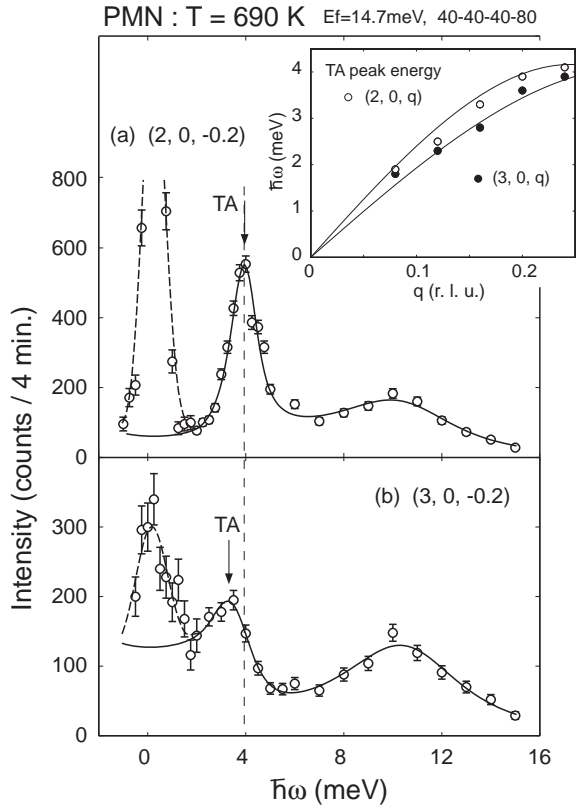}}
\caption{Constant-{\bf Q} scans measured at (a) {\bf Q} $ = (2, 0,
-0.2)$ and (b) {\bf Q} $ = (3, 0, -0.2)$ at 690\,K.  Solid lines
are fits to the mode-coupling function convoluted with the
instrumental resolution function.  A negative $F_{1}F_{2}$ is used
for $(2, 0, -0.2)$ while a positive $F_{1}F_{2}$ is used for $(3,
0, -0.2)$.  Although the scattering intensities from the TA modes
peak at different frequencies, the fits yield the same value for
$\Omega_{2} = 4.2$ (as is required), as well as the same value for
$\lambda = 17$.  Note that the TA structure factor $|F_{2}|^{2}$
at $(3,0,-0.2)$ is $\approx 100$ times smaller than it is at $(2,
0, -0.2)$. The inset shows the {\it apparent} phonon frequencies
of the TA modes determined solely from the peak position.  The TA
modes at $(3, 0, q)$ appear systematically at lower frequency than
do those at $(2, 0, q)$.}
\end{figure}

Another important difference between the (200) and (300) zones is
that the dynamic structure factor $|F_{2}|$ at $(3,0,-0.2)$ is
about ten times smaller than that for $(2,0,-0.2)$.  This fact is
readily confirmed since, for wave vectors close to the zone
center, the ratio of $|F_{2}|$ in two different zones should agree
with that for $|F_N|$, the nuclear static structure factor.  This
is given by
\begin{equation}
F_{N}({\rm\bf Q}) = \sum_{j} b_{j}e^{i{\rm\bf Q}\cdot{\rm\bf d}_{j}}e^{-W_{j}},
\end{equation}
where $b_{j}$, ${\rm\bf d}_{j}$, and $e^{-W_{j}}$ are the neutron
scattering length, atomic coordination vector, and the
Debye-Waller factor for the $j$th atom, respectively.  In fact,
this formula gives $|F_{N}(200)|/|F_{N}(300)| = 11.2$ which is in
excellent agreement with the value
$|F_{2}(2,0,-0.2)|/|F_{2}(3,0,-0.2)| = 12.0$ obtained from the
fits.  This implies, however, that the actual intensity of the TA
mode at $(3,0,-0.2)$ (which is proportional to $|F_2|^2$) should
be over 100 times smaller than that at $(2, 0, -0.2)$.  Yet the TA
peak is clearly visible at $(3, 0, -0.2)$.  This can also be
explained by the MC function, since the intensity of the TO mode
is transferred to the TA mode due to the coupling, which in turn
produces an enhancement of the TA cross section.  This feature is also
described in more detail in the Appendix.

Features similar to those discussed above are also found in the
phonon profiles at smaller $q$.  Data taken at $(3,0,-0.12)$ and
$(3,0,-0.08)$ are shown in Fig.~4(a) and (b), along with the fits
to the MC function.  As before, the solid and dashed lines
correspond to MC and Gaussian functions, respectively. The fitting
parameters are listed in Table~1. The coupling constant $\lambda$
was determined such that $\Omega_{1}$ and $\Omega_{2}$ are equal
in each zone.  At smaller $q$, the parameter $\Gamma_{1}$ becomes
larger, meaning that the TO mode is more heavily damped.  This is
consistent with the ``waterfall'' picture.  We have thus shown
that the phonon profiles in both the (200) and (300) Brillouin
zones can be explained satisfactorily by the MC function at all
$q$. However there is still an open question.  While we know that
$\lambda$ decreases at smaller $q$, we do not yet have a specific
model that describes the $q$-dependence of $\lambda$.

\begin{figure}
\centerline{\epsfxsize=3in\epsfbox{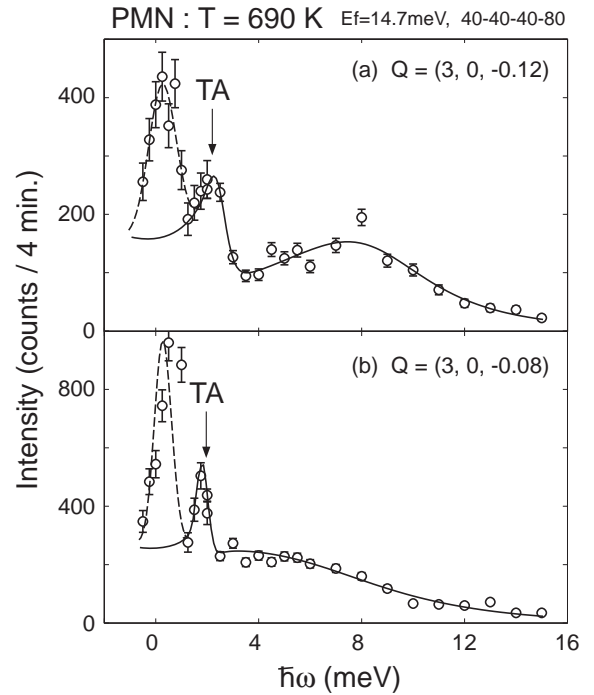}}
\caption{Constant-{\bf Q} scans measured at (a) {\bf Q} $= (3, 0,
-0.12)$ and (b) {\bf Q} $= (3, 0, -0.08)$.  Solid lines are fits
to the mode-coupling function convoluted with the instrumental
resolution function. Here $F_{1}F_{2}$ was chosen to be positive
for both scans. All parameter values are listed in Table~1.}
\end{figure}

\begin{table}
  \caption{Parameters obtained from the fit of the 690\,K data to
  the MC function.}
\begin{ruledtabular}
\begin{tabular}{lccccccc}
${\rm\bf Q}$
&
$\Omega_{1}$
&
$\Omega_{2}$
&
$\Gamma_{1}$
&
$\Gamma_{2}$
&
$|F_{1}|$
&
$|F_{2}|$
&
$\lambda$ \\
\hline
$(2, 0, -0.2)$
&
$10.85$
&
$4.30$
&
$6.10$
&
$0.79$
&
$27.0$
&
$13.1$
&
$17.4$ \\
$(3, 0, -0.2)$
&
$10.92$
&
$4.20$
&
$6.05$
&
$1.95$
&
$27.0$
&
$1.09$
&
$17.4$ \\
\hline
$(3, 0, -0.12)$
&
$9.05$
&
$2.67$
&
$7.77$
&
$0.69$
&
$25.7$
&
$<0.5$
&
$7.21$ \\
$(3, 0, -0.08)$
&
$8.76$
&
$1.92$
&
$13.1$
&
$0.03$
&
$31.3$
&
$0.64$
&
$1.88$ \\
\end{tabular}
\end{ruledtabular}
\end{table}

\subsection{Temperature dependence of the phonon cross section}

To check the consistency of our model, we have analyzed the TA
phonon profiles at $(3,0,-0.2)$ at different temperatures.  Before
showing the results for $(3,0,-0.2)$, we first present the
temperature dependence of the phonon profiles obtained at
$(2,0,-0.2)$.  These are displayed in the left panel of Fig.~5 at
(a) 700\,K, (b) 500\,K, and (c) 290\,K.  Each profile has been fit
to a resolution-convoluted MC function plus a Gaussian centered at
$\hbar\omega = 0$ with $F_{1}F_{2} < 0$.  The fitted parameter
values are listed in Table~2.  Consistent with the results of the
previous section, $|F_{1}|/|F_{2}| \approx 2$ at 700\,K, and is
almost independent of temperature down to 290\,K.  Similarly,
$\Omega_{1}$, $\Omega_{2}$, and $\lambda$ are also independent of
temperature.  On the other hand, $\Gamma_{2}$ (the TA mode
linewidth) is larger at 500\,K and 290\,K, which is consistent
with the TA linewidth broadening below $T_{d}$ reported in
Ref.~7 The linewidth of the TO mode $\Gamma_1$ is
also slightly broader below 700\,K.

The right panel of Fig.~5 shows the temperature dependence of the
TA phonon for {\bf Q} = (3,0,-0.2) measured at (d) 570\,K, (e)
350\,K, and (f) 50\,K.  Because these data do not include the TO
mode, the profiles for 570\,K and 350\,K were fit fixing
$\Omega_{1}=11.2$, $\Gamma_{1}=6$, and $\lambda=17.4$, while
$F_{1}F_{2} > 0$.  Consequently, the uncertainties associated with
the parameters in this zone will be larger.  Values for the fitted
parameters are listed in Table~2.  From this analysis we find that
the ratio $|F_{1}(200)|/|F_{1}(300)| \approx 1$.  Since the
nuclear structure factor calculations imply
$|F_{2}(200)|/|F_{2}(300)| \approx 10$, and we know that
$|F_{1}(200)|/|F_{2}(200)| \approx 2$, we can estimate that
$|F_{1}(300)|/|F_{2}(300)| \approx 20$.  This agrees reasonably
well with the fitted ratio $|F_{1}|/|F_{2}| \approx 16$ at both
temperatures, and represents a reassuring self consistency check
of our model.  Moreover, a broadening of the TA mode is found at
(3,0,-0.2) that is qualitatively similar to that observed at
(2,0,-0.2).  Thus, the temperature dependence of the phonon
profiles is consistent with the mode-coupling model.  However we
did not obtain a clear TA phonon peak at 50\,K.  Instead, the
solid line shown in Fig.~5(f) represents the profile that would be
expected for the TA mode assuming the same parameters as those
obtained from fits to the 570\,K data.  The agreement between the
data and fit in this case is quite good.

\begin{table}
  \caption{Parameters obtained from the fit of the data in Fig.~5 to
  the MC function.}
\begin{ruledtabular}
\begin{tabular}{lcccccccc}
${\rm\bf Q}$
&
$T (K)$
&
$\Omega_{1}$
&
$\Omega_{2}$
&
$\Gamma_{1}$
&
$\Gamma_{2}$
&
$|F_{1}|$
&
$|F_{2}|$
&
$\lambda$ \\
\hline
$(2, 0, -0.2)$
&
$700$
&
$11.1$
&
$4.60$
&
$5.24$
&
$1.24$
&
$10.5$
&
$5.65$
&
$17.4$ \\
$ $
&
$500$
&
$11.1$
&
$4.45$
&
$6.07$
&
$1.70$
&
$12.5$
&
$6.07$
&
$16.8$ \\
$ $
&
$290$
&
$12.0$
&
$4.90$
&
$5.96$
&
$1.52$
&
$11.5$
&
$6.04$
&
$17.9$ \\
\hline
$(3, 0, -0.2)$
&
$570$
&
$11.2$
&
$3.90$
&
$6.00$
&
$2.89$
&
$12.1$
&
$0.78$
&
$17.4$ \\
$ $
&
$350$
&
$11.2$
&
$4.29$
&
$6.00$
&
$4.81$
&
$14.0$
&
$0.82$
&
$17.4$ \\
\end{tabular}
\end{ruledtabular}
\end{table}

\begin{figure}
\centerline{\epsfxsize=3.4in\epsfbox{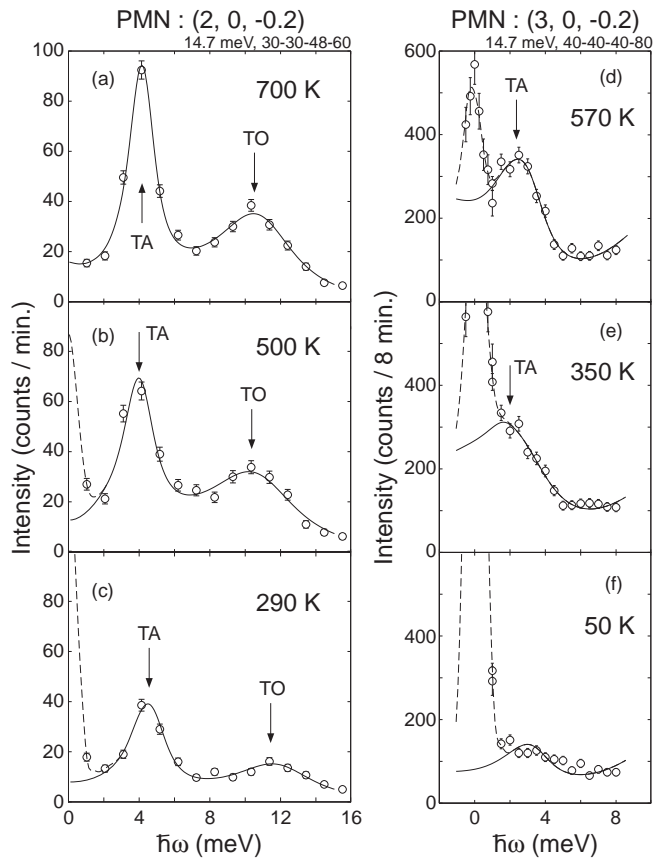}}
\caption{The left panel shows phonon profiles measured at {\bf Q} 
$= (2, 0, -0.2)$ at (a) 700\,K, (b) 500\,K, and (c) 290\,K.  The 
right panel shows TA phonon profiles measured at {\bf Q} = 
$(3, 0, -0.2)$ at (d) 570\,K, (e) 350\,K, and (f) 50\,K.  
The lines, except those shown in (f), represent fits to the 
resolution-convoluted MC function plus a Gaussian function 
centered at $\hbar\omega=0$. The solid and dashed lines 
correspond to the MC and Gaussian cross sections, respectively. 
The solid line in (f) is an MC cross section using the same 
parameters as those obtained in (d).}
\end{figure}

\subsection{Dynamic structure factor for TO cross section}

The ratio of the dynamic structure factors that govern the 
transverse optic phonon intensities in different Brillouin 
zones depends on the associated atomic vibrational displacements.
Two of the most important atomic vibrational modes are those 
proposed by Slater and Last, and are discussed by Harada 
{\it et al.}~\cite{Harada_70}Ê The Slater mode corresponds to 
atomic motions in which the oxygen and Mg/Nb atoms move in 
opposition while the Pb atoms remain stationary.Ê The Last mode 
corresponds to opposing motions of the Pb atoms and rigid 
(Mg/Nb)O$_{6}$ octahedra.Ê The dynamic structure factor is given by 
\begin{equation} 
\label{f_inel} 
F_{inel} = 
\sum_{j} [{\rm\bf Q}\cdot \xi_{j}] b_{j}e^{i{\rm\bf G}\cdot{\rm\bf d}_{j}}e^{-W_{j}}, 
\end{equation} 
where $\xi_{j}$ is the normalized displacement vector for the $j$th 
atom.Ê Assuming only the Slater and Last modes are significant, then 
$\xi_{j}$ can be written in terms of the atomic displacement vectors 
for the Slater (${\rm\bf s}_{j1}$) and Last (${\rm\bf s}_{j2}$) modes,
and the parameter $S$, which is a measure of the relative contributions 
of the Last and Slater modes, as follows: 
\begin{equation} 
\xi_{j} = {\rm\bf s}_{j1} + S {\rm\bf s}_{j2}. 
\end{equation} 
For PMN, Hirota {\it et al.}~\cite{Hirota_01} calculated the value 
of $|F_{inel}|^2$ for $-0.5 \leq S \leq 2$ at (200), (300), and (110) 
as shown in Fig.~6 of Ref.~11. Our result of 
$|F_{1}(200)|/|F_{1}(300)| \approx 1$ obtained from our fits to the MC 
function is consistent with the value $S=1.5$ determined by Hirota 
{\it et al.}Ê As an additional check, we measured the TO cross section 
at 80\,K at several zone centers in the [hhl] zone.Ê Because the TO 
cross section at 80\,K is greatly diminished by the Bose factor, we 
were unable to perform a quantitative comparison of the phonon 
intensities to the dynamic structure factors using Eq.~\ref{f_inel}.Ê 
However, we find that the TO phonon cross section at $(111)$ is quite 
small.Ê This observation agrees qualitatively with the calculations 
of Hirota {\it et al.}~\cite{Hirota_01} using $S=1.5$ for which 
$|F_{inel}(111)|^2/|F_{inel}(200)|^2 = 0.14$.

\section{Discussion}

In the previous section we presented our mode-coupling analysis
and demonstrated that the characteristic features of the phonon
profiles in PMN can be explained by a coupling between the TO and
TA modes, without need for any additional mode. This fact implies
that the TO mode we observe is the ferroelectric soft mode.
However our interpretation conflicts with the model proposed by
Naberezhnov {\it et al.},~\cite{Naberezhnov_99} and later refined
by Vakhrushev and Shapiro,~\cite{Vakhrushev_02} which requires an
additional soft mode.  To reconcile these two models, we made a
careful comparison of their data with ours.  We find that, where
there is overlap, the data agree quite well, and that only the
interpretations are different.  In this section, we attempt to
explain the possible reasons for this difference.

\subsection{Naberezhnov-Vakhrushev branch}

In 1999 Naberezhnov {\it et al.} published an extensive study of
the lattice dynamics of PMN.~\cite{Naberezhnov_99}Ê They observed
well-defined TA and low-lying TO modes in the vicinity of the (220) 
zone center at 800\,K.Ê Surprisingly, they found that the structure 
factors of the TO modes are inconsistent with those derived from 
the diffuse scattering intensities reported earlier by Vakhrushev 
{\it et al.}~\cite{Vakhrushev_89,Vakhrushev_93}Ê This conclusion 
was based on the fact that the normalized displacements for the 
Pb ($\delta(Pb) = 1.00$), Mg or Nb ($\delta(MN) = 0.18$), and 
oxygen atoms ($\delta(O) = -0.64$), determined by Vakhrushev 
{\it et al}.,~\cite{Vakhrushev_93} do not satisfy the 
center-of-mass condition.Ê Naberezhnov {\it et al}.\ noted that the 
center of mass of the unit cell is shifted from its original 
position, and thus cannot correspond to optical lattice vibrations.Ê
They attributed this shift to the slow relaxation of superparaelectric
clusters.Ê They therefore concluded that the observed low-lying TO
mode could not be the ferroelectric soft mode.Ê Instead they 
identified it as a hard TO1 mode.Ê A ``quasi-optic'' (QO) was mode was 
derived by fitting the TA profiles at the (200) and (110) zones to a 
function that couples the TA and QO modes, and this QO mode was 
identified as the soft mode.Ê However, the inconsistency between the 
structure factors associated with the diffuse scattering and the TO 
mode has since been resolved by the ``phase-shifted condensed soft 
mode'' model proposed by Hirota {\it et al.}~\cite{Hirota_01}Ê Hirota 
{\it et al}.\ realized that the atomic displacements can be separated 
into two components: a uniform phase shift $\delta_{shift} = 0.58$, 
and displacements that do satisfy the center-of-mass condition, namely 
$\delta_{cm}(Pb) = 0.42$, $\delta_{cm}(MN) = -0.40$, and 
$\delta_{cm}(O) = -1.22$.Ê Moreover, these center-of-mass displacements
are consistent with the observed phonon cross section in that they 
correspond to a TO mode that is a mixture of Slater and Last modes in 
the ratio $S=1.5$.Ê Thus, the ``phase-shifted condensed soft mode'' 
model is able to reconcile the intensities of the TO phonon and diffuse 
scattering, and demonstrates that the diffuse scattering originates 
from the condensation of the soft TO mode. 

Vakhrushev and Shapiro have since analyzed the phonon profiles in
the (300) zone where the TA phonon cross section is expected to be
extremely weak as a result of a small dynamic structure factor.
Their analysis uses a function that couples the TA and QO modes
and neglects the TA cross section because it was expected to be
weak.~\cite{VS_neglect}  But as demonstrated in Fig.~3(b) and in
the appendix, the coupling between modes greatly enhances the TA
phonon cross section in spite of the small structure factor.  Thus
the TA mode is not negligible. The data taken by Vakhrushev and
Shapiro~\cite{acknowledge_VS} shows peaks that are completely
identical to the TA peaks we observe. For example, they observe a
peak at $\approx 2$\,meV at (3,0.075,0) at 880\,K (Fig.~3 in
Ref.~17), which is completely consistent with
the TA peak we observe in Fig.~4(b).  Another data of the identical 
sample at $(3, -0.1, 0)$ at 880~K is also published in Fig. 3 of 
Ref.~4.  However, these peaks were
treated as a ``Bragg tail''~\cite{Gens_book} in the analyses of 
Ref.~17.  Again, the TA cross
section is enhanced by the coupling and therefore one cannot
neglect the TA cross section even though the structure factor for
the TA mode in the (300) Brillouin zone is very small. Our model
of coupling between TA and TO modes describes all of the
characteristics of the phonon profiles and thus removes the need
for an additional QO branch.

\begin{figure}
\centerline{\epsfxsize=3in\epsfbox{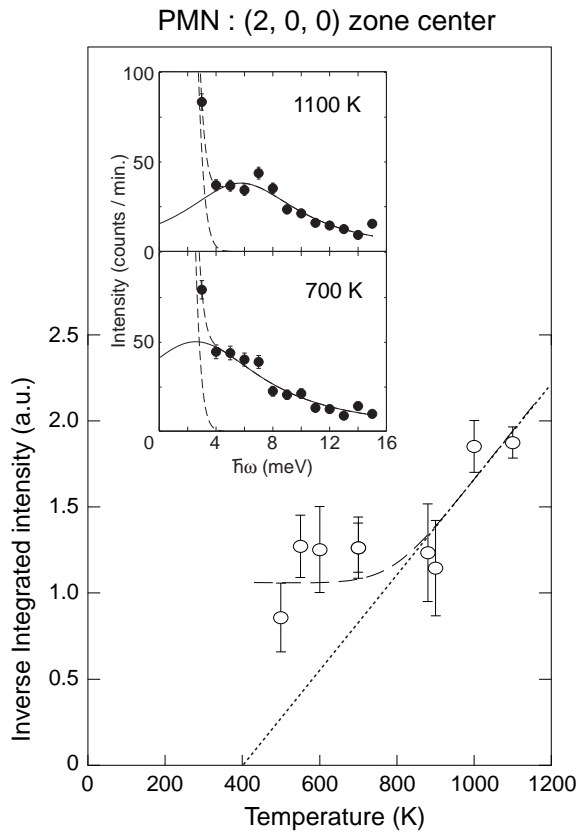}}
\caption{Temperature dependence of the inverse integrated
intensity of the zone center TO phonon.  The inset shows TO phonon
profiles measured at 1100\,K and 700\,K.  The solid and dashed
lines correspond to a damped harmonic oscillator description of
the TO mode and a Gaussian form centered on $\hbar\omega_{0}=0$,
respectively (see text).  The inverse integrated intensities are
calculated from a numerical summation of the TO spectral weight
corrected for the Bose factor. The high temperature data ($T >
800$\,K) agree with the Curie-Weiss behavior measured by Viehland
{\it et al}., shown by the dotted line, for which $1/\epsilon
\propto (T-T_{0})$, where $T_0 = 400$\,K.}
\end{figure}

\subsection{Dielectric constant}

Because the QO mode is an artifact of the mode-coupling cross
section, the Curie-Weiss behavior of the static dielectric
constant must come from the softening of the zone-center TO mode.
Therefore this TO mode is the ferroelectric soft mode.  In fact,
the zone-center TO mode measured at (200) exhibits a significant
temperature dependence as shown in Fig.~1, and the linear
temperature dependence of $(\hbar\omega_0)^2$ shown in Fig.~1(b)
is qualitatively similar to that observed in conventional
soft-mode ferroelectrics, such as PbTiO$_3$.~\cite{Shirane_70}
However, the temperature dependence of $(\hbar\omega_0)^2$ in
Fig.~1(b) does not satisfy the Lyddane-Sachs-Teller (LST) relation
\begin{equation}
\label{LST} 1/\epsilon \propto (\hbar\omega_{0})^{2},
\end{equation}
where $\epsilon$ represents the static dielectric constant.
Measurements of $\epsilon$ by Viehland {\it et
al}.~\cite{Viehland_92} for $T \ge 600$\,K show that the static
dielectric constant exhibits a Curie-Weiss behavior
\begin{equation}
\label{CW} \epsilon \propto 1/(T-T_0)\, ,
\end{equation}
with $T_0 = 400$\,K, which is shown by the dotted line in
Fig.~1(b).  On the other hand, a linear extrapolation of
$(\hbar\omega_0)^2$ for $T > T_d$ gives $T_0 \approx 0$\,K. This
was pointed out by Vakhrushev and Shapiro in
Ref.~17.  We have since reanalyzed the data and
confirmed that the temperature dependence of $(\hbar\omega_0)^2$
indeed follows the solid line shown in Fig.~1(b).

Vakhrushev and Shapiro found that $T/I(T)$ for $T>T_{d}$ varies as
$T-T_{0}$ with $T_{0}=340$\,K, where $I(T)$ is the integrated
intensity measured at $(3,0,0)$ over the energy range $2 <
\hbar\omega < 12$\,meV.  We have performed a similar analysis on
our data measured at the (200) zone center which shows
well-defined TO peaks.  We fit the (200) zone center phonon
profiles to a resolution-convoluted cross section composed of a
Gaussian component to describe the elastic ($\hbar\omega = 0$)
scattering plus a damped harmonic oscillator function to describe
the TO mode.  The fits to the data at 1100\,K and 700\,K are shown
in the insets to Fig.~6. The solid lines represent the damped
harmonic oscillator component.  We next evaluated the Bose-factor
corrected integrated intensity for the damped harmonic oscillator
component by numerical summation.  Figure~6 shows the temperature
dependence of the inverse integrated intensity so obtained. The
high temperature data agree with the Curie-Weiss behavior of
$1/\epsilon$ reported by Viehland {\it et al.}, with $T_0 =
400$\,K,  which is shown as a dotted line in
Fig.~6.

\begin{figure}
\centerline{\epsfxsize=3in\epsfbox{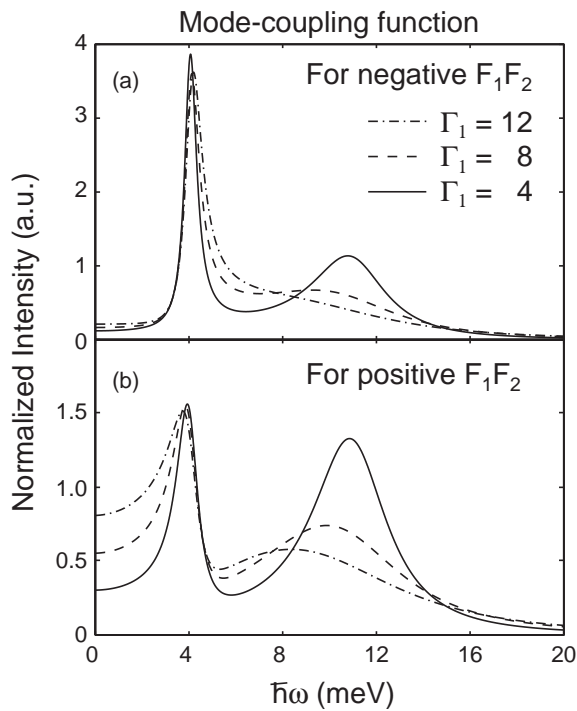}}
\caption{Mode-coupling profiles as a function of $\Gamma_{1}$ with
(a) $F_{1}F_{2} = -2$ (negative), and (b) $F_{1}F_{2} = 85$
(positive).  The other parameters were held fixed at the values
$\Omega_{1}=11, \Omega_{2}=4, \Gamma_{2}=1$, and $\lambda=20$,
which correspond to those obtained for $(2, 0, -0.2)$ and $(3, 0,
-0.2)$.  For negative $F_{1}F_{2}$, the spectral weight of the
damped TO mode shifts toward the region between the TA and TO
modes, and causes the apparent TA peak position to move to higher
frequency.  On the other hand, for positive $F_{1}F_{2}$, the
spectral weight tends to shift towards the elastic
($\hbar\omega=0$) position, and this causes the apparent TA peak
position to move to lower frequency.}
\end{figure}

\subsection{Concluding remarks}

We have established the coupled nature of the TA and TO modes in
PMN at 690\,K, which lies above the Burns temperature $T_{d}$.
This coupling, along with the soft character of the TO mode, leads
naturally to the idea of a soft {\it coupled optic mode} that
condenses at $T_d$, and which therefore contains a significant
transverse acoustic component.  This idea was originally suggested
by Y.\ Yamada.~\cite{Yamada_model}  
The condensation of such a coupled optic mode
would then provide an elegant explanation for the origin of the
phase-shifted condensed soft mode model of the PNR proposed by
Hirota {\it et al.}~\cite{Hirota_01}  If correct, this concept may
be the key to understanding the fundamental mechanism underlying
the unusual relaxor behavior.  We have not as yet established a
definite connection between such a soft coupled TO mode and the
phase-shifted condensed soft mode.  However this is one of our
future projects.

\begin{acknowledgments}
Stimulating conversations with Y.\ Yamada provided us with one
of the key concepts presented in this paper.Ê We also gratefully 
acknowledge S.\ B.\ Vakhrushev and S.\ M.\ Shapiro for sharing their 
data prior to publication, and we thank R.\ J.\ Birgeneau, A.\ A.\ 
Bokov, K.\ Hirota, and J.\ -M.\ Kiat for interesting discussions.Ê 
Work at the University of Toronto is supported by the Natural Science 
and Engineering Research Council of Canada.Ê Finally, we acknowledge 
financial support from the U.\ S.\ DOE under contract No.\ 
DE-AC02-98CH10886, and the Office of Naval Research under Grant No.\ 
N00014-99-1-0738. 
\end{acknowledgments}

\begin{figure}
\centerline{\epsfxsize=3in\epsfbox{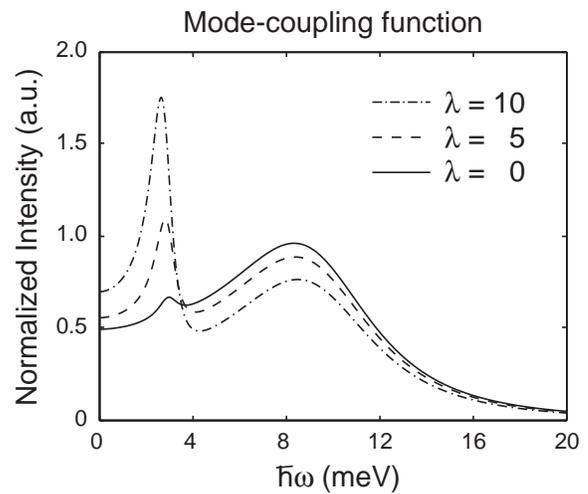}}
\caption{Mode-coupling profiles as a function of $\lambda$. Values
for the other parameters were held fixed at $\Omega_{1}=10,
\Omega_{2}=3, \Gamma_{1}=8, \Gamma_{2}=1$, and $F_{1}/F_{2}=24$,
which correspond approximately to those obtained at {\bf Q} = $(3,
0, -0.12)$ in Fig.~4(a).  Part of the TO phonon intensity is shown
to transfer to the TA mode as a result of the coupling, and
results in a substantial enhancement of the TA mode intensity.}
\end{figure}

\section{Appendix}

The phonon cross sections presented here have all been fit using
the MC function (Eqs. 2, 3, and 4) given by Harada {\it et
al.}~\cite{Harada_71} In this appendix we present several
important features of the MC function that give rise to large
differences in the phonon profiles measured in different Brillouin
zones.

The effects of coupling can vary substantially when the sign of
$F_{1}F_{2}$ is changed.  Figure~7 shows the cross sections
calculated using the MC function. The product of the two dynamic
structure factors $F_{1}F_{2}$ was set to be $-2$ (negative) for
Fig.~7(a), and $85$ (positive) for Fig.~7(b). In both figures
three cross sections are plotted for $\Gamma_{1} = 12$, $8$ and
$4$, while the other parameters are fixed at $T = 690$\,K,
$\Omega_{1} = 11$\,meV, $\Omega_{2} = 4$\,meV, $\Gamma_{2} =
1$\,meV, and $\lambda = 20$. These parameters are approximately
the same as those for obtained from fits to data measured for {\bf
Q} $= (2,0,-0.2)$ and $(3,0,-0.2)$.  The intensities have been
normalized so that the integrated cross section is kept constant.
The TO phonon cross section for negative $F_{1}F_{2}$ shifts
increasingly to the energy region between the TA and TO modes with
increasing $\Gamma_{1}$.  On the other hand, the TO phonon
spectral weight shifts to the energy region below the TA mode near
$\hbar\omega = 0$ for positive $F_{1}F_{2}$.  Because of this, the
apparent TA peak position tends to shift to higher energy for
negative $F_{1}F_{2}$, and to lower energy for positive
$F_{1}F_{2}$.  Thus, a change in the sign of $F_{1}F_{2}$ causes
the apparent TA peak position to shift in different directions.

Another important effect of the mode-coupling is shown in Fig.~8,
which displays MC cross sections for $\lambda = 0$, $5$, and $10$.
The other parameters are fixed at $T=690$\,K, $\Omega_{1}=10$,
$\Omega_{2}=3$, $\Gamma_{1}=8$, $\Gamma_{2}=1$, and
$F_{1}/F_{2}=24$. Again the cross sections have been normalized to
keep the integrated cross section constant.  As a result of the
coupling between modes, the TO cross section is partially
transferred to the TA cross section.  With sufficiently large
values of $\lambda$, a remarkably large enhancement of the TA peak
can occur.


\begin{references}

\bibitem{Ye_review} See review article, Z.-G. Ye, {\it Key Engineering
    Materials Vols. 155-156}, 81 (1998).
\bibitem{Naberezhnov_99} A. Naberezhnov, S. Vakhrushev, B. Doner, D.
  Strauch, and H.  Moudden Eur. Phys. J. B {\bf 11}, 13 (1999)
\bibitem{Gehring_00_1} P. M. Gehring, S.-E. Park, and G. Shirane,
  Phys. Rev. Lett. {\bf 84}, 5216 (2000).

\bibitem{Gehring_proc_00} P. M. Gehring, S. B. Vakhrushev, and G. Shirane,
  Proceeding of Fundamental Physics of Ferroelectrics,  Aspen 2000.
\bibitem{Gehring_00_2} P. M. Gehring, S.-E. Park, and G. Shirane,
  Phys. Rev. B {\bf 63}, 224109 (2000).
\bibitem{Gehring_01} P. M. Gehring, S. Wakimoto, Z.-G. Ye, and G.
  Shirane, Phys. Rev. Lett. {\bf 87}, 277601 (2001).
\bibitem{waki_02} S. Wakimoto, C. Stock, R. J. Birgeneau, Z.-G. Ye,
  W. Chen, W. J. L. Buyers, P. M. Gehring, and G. Shirane,
  Phys. Rev. B {\bf 65}, 172105 (2002).
\bibitem{Burns_83} G. Burns and F. H. Dacol, Solid State Commun. {\bf
    48}, 853 (1983).
\bibitem{Vakhrushev_89} S. B. Vakhrushev, B. E. Kvyatkovksy, A. A.
  Naberezhnov, N. M. Okuneva, and B. Toperverg, Ferroelectrics {\bf
    90}, 173 (1989).
\bibitem{Vakhrushev_93} S. B. Vakhrushev, A. A. Naberezhnov, N. M.
  Okuneva, and B. N. Savenko, Phys. Solid State {\bf 37} 1993 (1995).
\bibitem{Hirota_01} K. Hirota, Z.-G. Ye, S. Wakimoto, P. M. Gehring,
  and G.  Shirane, Phys. Rev. B {\bf 65}, 104105 (2002).
\bibitem{Harada_71} J. Harada, J. D. Axe and G. Shirane,
  Phys. Rev. B {\bf 4}, 155 (1971).
\bibitem{deMathan} N de Mathan, E. Husson, G. Calvarin, J. R.
  Gavarri, A.  W. Huwat, and A. Morell, J. Phys. Condens. Matter {\bf
    3}, 8159 (1991).
\bibitem{Tu_95} C.-S. Tu, V. Hugo Schmidt, and I. G. Siny, J. Appl.
  Phys. {\bf 78}, 5665 (1995).
\bibitem{Dkhil_02} B. Dkhil, J. M. Kiat, G. Calvarin, G. Baldinozzi,
  S. B.  Vakhrushev, and E. Suard, Phys. Rev. B {\bf 65}, 024104
  (2002).

\bibitem{Harada_70} J. Harada, J. D. Axe, and G. Shirane, Acta Cryst. 
  {\bf A 26}, 608 (1970).
\bibitem{Vakhrushev_02} S. B. Vakhrushev and S. M. Shapiro, unpublished
(cond-mat/0203103).
\bibitem{VS_neglect} This is explicitly stated in page 3 of Ref. 17.
\bibitem{acknowledge_VS} Vakhrushev and Shapiro kindly provided us with
data sets taken at several different temperatures.

\bibitem{Gens_book} For details about Bragg tails, see p.\,114 of {\it Neutron
scattering with a triple axis spectrometer} by G. Shirane, S. M.
Shapiro, and J. M. Tranquada, Cambridge University Press.  We
confirmed that the TA mode we observed is not a Bragg tail from
the fit to the MC function. The parameter $F_1$ is nearly
$q$-independent. If the peak were a Bragg tail, $F_1$ would depend
strongly on $q$.
\bibitem{Shirane_70} G. Shirane, J. D. Axe, and J. Harada,
  Phys. Rev. B {\bf 2}, 155 (1970).
\bibitem{Viehland_92} D. Viehland, S. J. Jang, L. E. Cross, and M. Wuttig,
  Phys. Rev. B {\bf 46}, 8003 (1992).
\bibitem{Yamada_model} Y. Yamada, private communication.


\end{references}
\end{document}